\documentclass{article}

\usepackage{times}
\usepackage{graphicx} 
\graphicspath{{figures/}} 

\usepackage{subfigure}

\usepackage{natbib}

\usepackage{algorithm}
\usepackage{algorithmic}
\usepackage{amsmath,amssymb,amsthm}

\usepackage{hyperref}

\usepackage[accepted]{icml2017}

\icmltitlerunning{Multifidelity Risk Estimation - Submitted to ICML 2017}

\begin{document}

\twocolumn[
\icmltitle{Estimating the risk associated with transportation \\
           technology using multifidelity simulation}

\icmlauthor{Erik J. Schlicht}{ejs@c2-g.com}
\icmladdress{Computational Cognition Group, C2-g \\
            3432 Denmark Avenue, Suite 610 \\St. Paul, MN 55123 USA}
\icmlauthor{Nichole Morris}{nlmorris@umn.edu}
\icmladdress{HumanFIRST Laboratory, University of Minnesota \\
            200 Transportation \& Safety Building \\ 511 Washington Ave. SE \\ University of Minnesota \\ Minneapolis, Minnesota, 55455 USA}

\icmlkeywords{Multifidelity, Simulation, Risk, Technology evaluation, Bayesian networks, Monte-Carlo}

\vskip 0.3in
]

\begin{abstract}
This paper provides a quantitative method for estimating the risk associated with candidate transportation technology, \emph{before it is developed and deployed.}  The proposed solution extends previous methods that rely exclusively on low-fidelity human-in-the-loop experimental data, or high-fidelity traffic data, by adopting a multifidelity approach that leverages data from both low- and high-fidelity sources. The multifidelity method overcomes limitations inherent to existing approaches by allowing a model to be trained inexpensively, while still assuring that its predictions generalize to the real-world. This allows for candidate technologies to be evaluated at the stage of conception, and enables a mechanism for only the safest and most effective technology to be developed and released.
\end{abstract}

\section{Introduction}
\label{intro}
The integration of technology into vehicles has become exceedingly common across the transportation industry.  However, it is often unclear how changes to technology will impact subsequent system safety until \emph{after} the technology has been developed, deployed and widely adopted.

This is primarily due to the fact that high-fidelity computer simulations are required to quantitatively estimate the risk associated with transportation technology \emph{prior} to deployment.  These simulations must include accurate quantitative models of all the components of the human interacting with the technology.  For example, in order to evaluate the risk associated with candidate collision-avoidance technology in aerospace systems, high-fidelity models of the airspace encounters, airframe dynamics, equipped sensors, collision-avoidance logic and pilot decisions are required \citep{Maki:10,Chryssanthacopoulos:11, Mueller:16}.  The latter is only afforded since commercial pilots are required to adhere to collision avoidance recommendations in the case of a near midair collision.  Therefore, the pilot can be quantitatively represented as a reaction time distribution.

\begin{figure}[ht]
\vskip 0.2in
\begin{center}
\centerline{\includegraphics[width=\columnwidth]{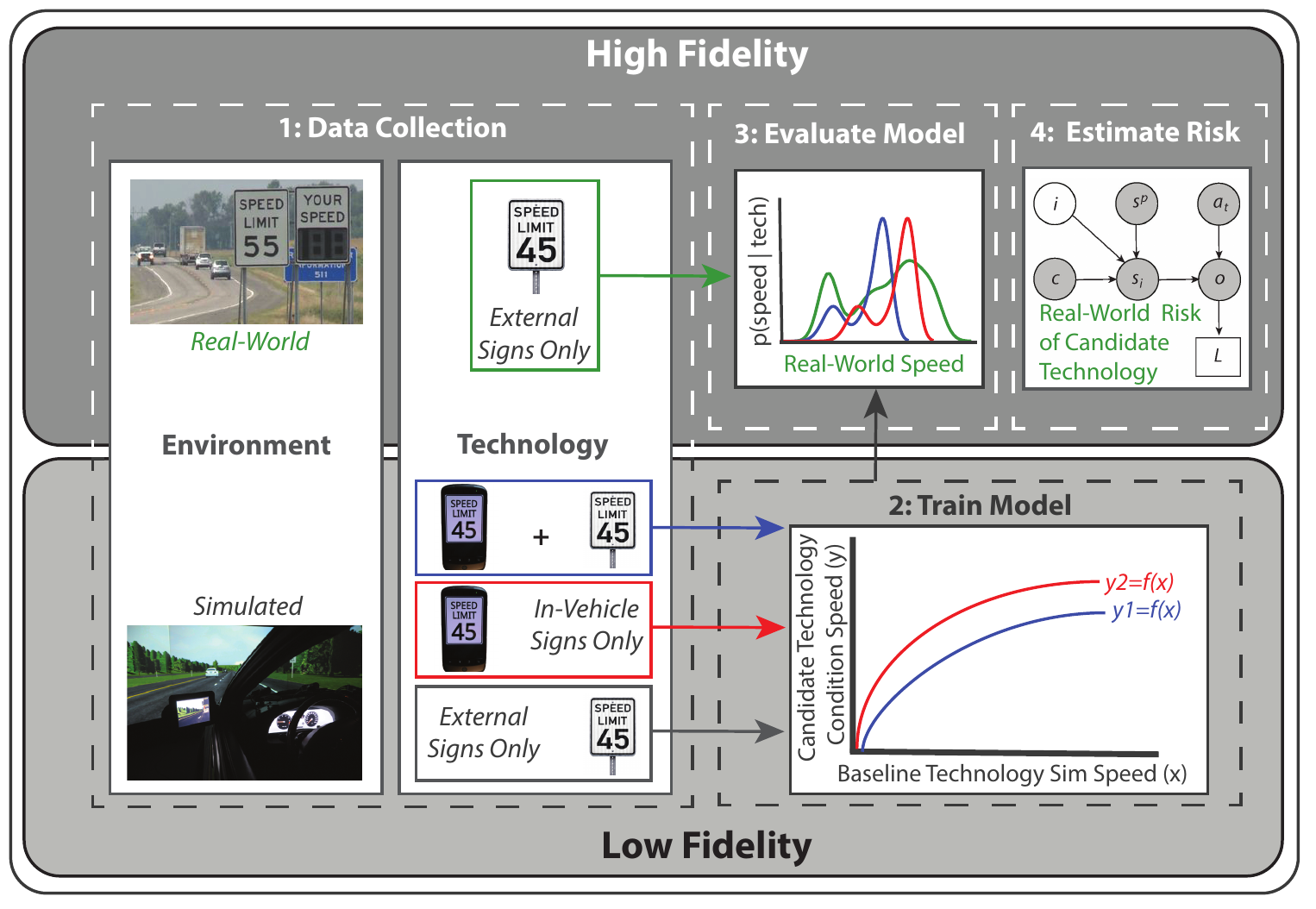}}
\caption{Diagram depicting the use of multifidelity methods for estimating the risk associated with transportation technology (i.e., In-vehicle sign (IVS) technology)}
\label{fig:multifidelity}
\end{center}
\vskip -0.2in
\end{figure}

In transportation systems where humans are not required to adhere to system recommendations, a quantitative model of how candidate technologies impact human decision-making is necessary.  However, the high-fidelity (i.e., real-world) data needed to develop these models is not available until after the technology has already been deployed. As a result, the ability to evaluate system safety has primarily focused on post hoc analyses that exclusively rely on high-fidelity data \cite{Yang:96,Agamennoni:12,Gindele:13,Lefevre:14,Morton:16,Wheeler:16}.  Therefore, in terms of time and development costs, these data are extremely expensive to obtain, and tend to provide only reactionary support to risks already imposed on a system.

Another approach is to rely on low-fidelity data that is obtained from human-in-the-loop driving simulation \cite{Gruening:98, Shechtman:09, Lee:03}.  Since the driving environment and technology can be experimentally manipulated, these data allow for insight into how driving behavior is impacted by changes to technology.  Moreover, since candidate technology can be evaluated prior to development, it saves time and money by allowing for changes to technology to be implemented before the development stage \cite{Gietelink:06}.

A major concern with using low-fidelity simulation data to estimate real-world risk, however, is that models developed using this data may not generalize to the real-world.  Therefore, exclusively relying on low-fidelity data to estimate safety is not a recommended practice, as it may lead to incorrect conclusions.

Fortunately, multifidelity approaches have been developed that leverage both low- and high-fidelity data in order to maximize the generalization of results, while minimizing the cost associated with estimation. Indeed, such approaches have been successfully used in wing-design optimization \cite{Robinson:2006}, robotic learning \cite{Cutler:15}, and have more recently been extended to human-in-the-loop systems \cite{Schlicht:2012}.

Previous work on multifidelity models of human-in-the-loop systems are theoretical and were used to predict pilot decisions during a self-separation encounter with an intruder aircraft \cite{Schlicht:2012}.  These theoretical efforts found that small differences between the simulated environment and the real-world do not impact the generalization of the model estimates.  More specifically, differences in simulation environment resulted in changes to the generative model (i.e,. Bayesian network) that reflected an absence of instrumentation in the low-fidelity (i.e., online) simulation environment.  These differences didn't lead to significant changes in the predicted action distributions between low-fidelity and high-fidelity environments \cite{Schlicht:2012}, provided participant expertise was held equivalent.

However, the same theoretical results found that participant expertise can adversely impact model predictions if there are significant differences in the actions taken by the real-world human actors (e.g., expert pilots) and the actions taken by experimental participants (e.g., novice pilots).  More specifically, the low-fidelity data should not be used to predict real-world performance if there is a significant difference in K-L Divergence between actions taken by experts and those taken by novices. This is due to the fact that models trained using low-fidelity data will not predict high-fidelity behavior \cite{Schlicht:2012}.

This study seeks to extend these theoretical findings \cite{Schlicht:2012} by utilizing a multifidelity approach to estimate the risk associated with candidate vehicle technology.  More specifically, we investigate the risk associated with in-vehicle signage (IVS) technology by leveraging a multifidelity predictive model in Monte-Carlo simulation (Figure \ref{fig:multifidelity}).  This method allows the risk of candidate transportation technology to be estimated \emph{prior} to being deployed, and is the first to use a multifidelity approach for risk estimation. The next section overviews the details of the low- and high-fidelity data which were collected for this effort.

\section{Low- and High-Fidelity Data}
In order to accurately estimate the risk associated with candidate transportation technology, it is desirable that the baseline condition in the low-fidelity simulation overlaps with real-world conditions (i.e., participant, environmental, and technological factors). This helps assure that the model developed using low-fidelity data generalizes well to the high-fidelity context. More specifically, it allows us to train a model that predicts driving performance (i.e., speed) under candidate technology conditions using only low-fidelity simulation data, once the similarity between performance is quantitatively verified.

Once the model is trained, we can use high-fidelity data as inputs to the model in order to \emph{predict} real-world driving behavior (i.e., speed) under technology conditions that have not yet been deployed. Finally, the predicted speed distributions can be used in a Monte-Carlo simulation to estimate the risk associated with the new technology. The next sub-section details how the low-fidelity data needed to train our models was collected.

\subsection{Low-Fidelity Data}
In the low-fidelity data collection study, baseline driving conditions (i.e., In-vehicle signage (IVS) technology absent and roadside signs available) were included as a control, in order to compare changes in relative safety and driving performance when using the IVS information. This low-fidelity condition was selected since it reflects the technology available to drivers in high-fidelity (i.e, real-world) settings. Therefore, by comparing a baseline condition to the two IVS conditions, one in which IVS information appears in conjunction with external (i.e., roadside) signs (IVS +ES) and one in which IVS information presented in replacement of external sign information (IVS -ES), it was expected that we would be able to identify any safety and workload effects that may be associated with the IVS technology.

More specifically, a 2 (Technology Condition: IVS +ES, IVS -ES) x 2 (IVS Condition: IVS absent (baseline - only roadside signage), IVS present) mixed-subjects factorial design was utilized, where subjects were randomly assigned to a technology condition (between-subjects factor), but participated in each IVS condition (within-subjects factor).  Notice that this design allows us to check for both main effects and interactions associated with the IVS technology.

\begin{figure}[ht]
\vskip 0.2in
\begin{center}
\centerline{\includegraphics[width=\columnwidth]{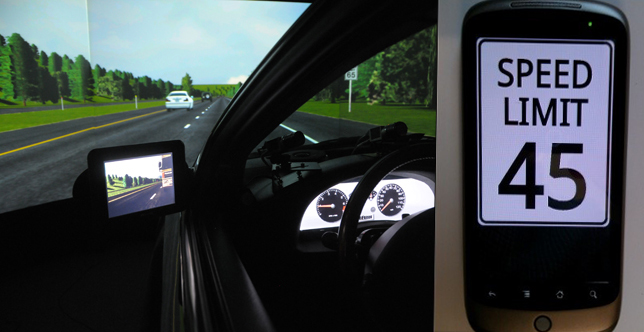}}
\caption{Low-fidelity data collection environment and simulated IVS technology}
\label{fig:simulator}
\end{center}
\vskip -0.2in
\end{figure}

This study was conducted in a partial motion-base driving simulator manufactured by Realtime Technologies (RTI). The simulator consisted of a 2002 Saturn SC2 full vehicle cab featuring realistic control operation and instrumentation including power-assist for the brakes and force feedback for the steering. Haptic feedback was provided by car body vibration and a three-axis electric motion system producing roll, pitch and yaw motion within a limited range of movement. The auditory feedback was provided by a 3D surround sound system. The driving environment was projected to a five-channel, 210-degree forward visual field screen (2.5 arc-minutes per pixel) with rear and side mirror views provided by a rear screen and vehicle-mounted LCD panels, respectively.
(Figure \ref{fig:simulator}).

IVS information was displayed to drivers on an Android cellular phone that was mounted to the center console of the vehicle within the drivers field of view. The description of each zone type, the speed limit for the zone. IVS information was presented visually only with no accompanying auditory alert or verbal information. The low-fidelity data collection methods and participants are described in detail in a MnDOT technical report \cite{Schlicht:2016}.

\subsection{High-Fidelity Data}
The high-fidelity speed data that was provided by the Minnesota Department of Transportation (MnDOT) for this effort. Participants included motorists who were on the road during MnDOT data collection, which occurred across speed zones during year 2014.  We have no reason to believe that the participants sampled from this data collection effort differed significantly from those sampled during our low-fidelity data collection \cite{Schlicht:2016}.

The roadways used during this high-fidelity data collection were those which have at least 1000 Annual Average Daily Traffic (AADT).  The speed of 100 vehicles were measured in each direction, and an attempt was made to ensure the sampling was done on a clear and sunny day during the MnDOT speed assessment. Moreover, samples were taken from an unmarked car, and they only sampled vehicles that are driving under free flow conditions.

\subsection{Comparison of Empirical Speed Data}

\begin{figure}[ht]
\vskip 0.2in
\begin{center}
\centerline{\includegraphics[width=\columnwidth]{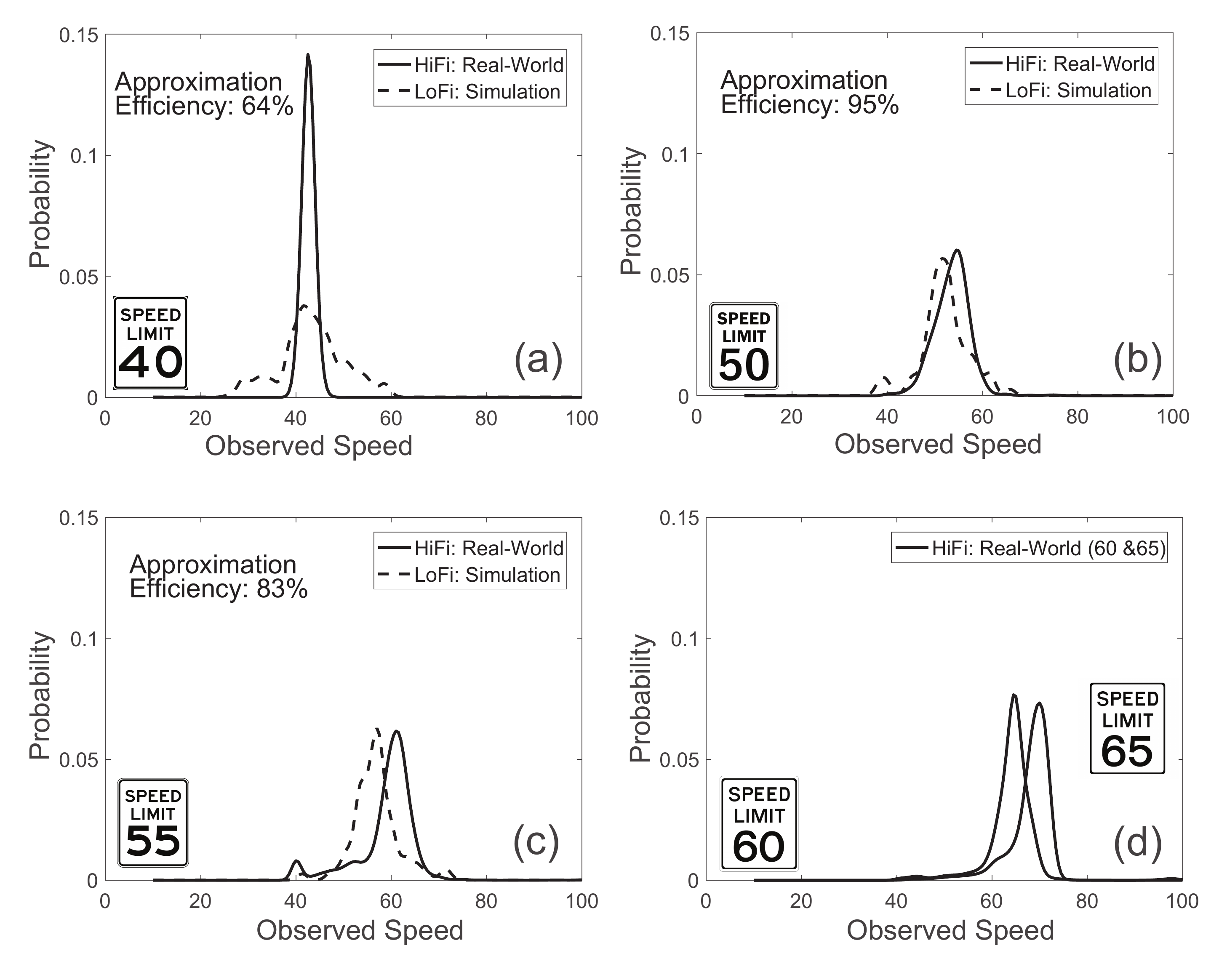}}
\caption{Comparison of low- and high-fidelity empirical speed distributions}
\label{fig:speeddists}
\end{center}
\vskip -0.2in
\end{figure}

A comparison between our high-fidelity speed data and the observed low-fidelity simulation data is depicted in Figure \ref{fig:speeddists}.  Kernel density estimation (with kernel width of 1) was performed on both sets of speed data in order to estimate $p(s \mid s^p,c = $ baseline technology$)$, which is the probability of the observed speed ($s$), given the posted speed ($s^p$) and (external-signs only) technology condition ($c$).

Figure \ref{fig:speeddists}a-d shows the high-fidelity speed distributions for each of the speed zones that were sampled in our high-fidelity data collection (40, 50, 55, 60, and 65mph). The low-fidelity data depicted are only for the simulation baseline conditions that overlap with our high-fidelity speed zones (Figure \ref{fig:speeddists}a-c).

It is apparent that there is a great degree of overlap between our observed low- and high-fidelity speed distributions across similar posted speed zones and technology conditions. The degree of overlap can be qualitatively measured using K-L Divergence ($\mathbb{K}(P||Q) = \sum_i log_2(p_i/q_i)p_i$), where the $P = p_{HiFi}(s \mid s^p,c = $ baseline technology$)$ distribution is defined to be the true distribution (i.e., high-fidelity speed distribution), that we approximate by using the $Q = p_{LoFi}(s \mid s^p,c = $ baseline technology$)$ distribution (i.e., low-fidelity speed distribution).

In Information Theory, K-L Divergence is a measure of information gain due to the use of an approximation to the true distribution, rather than the distribution itself \cite{Cover:06}. Since we know the true (high-fidelity) distribution ($P$) of the observed speeds, it is possible to describe the distribution with an average description length equal to the Shannon Entropy ($H(P) = -\sum_i p_ilog_2(p_i)$). However, if we instead used the low-fidelity distribution ($Q$) to approximate $P$, we would need $H(P)+\mathbb{K}(P||Q)$ bits, on average, to describe high-fidelity speed. In other words, we need $I(Q) = \mathbb{K}(P||Q)/H(P) \times 100\%$ more bits of information if we were to use the low-fidelity speed data ($Q$) as an approximation to the high-fidelity distribution ($P$). We refer to this quantity ($E(Q) = 100\% - I(Q)$) the \emph{approximation efficiency} and it is provided for the relevant conditions in Figure \ref{fig:speeddists}a-c.

As the $E(Q)$ metric suggests, the low-fidelity data provide a reasonably efficient approximation to the real-world distributions (mean $E(Q)$ = $81\%$ efficiency). This result agrees with previous theoretical work \cite{Schlicht:2012} that demonstrated small differences between low- and high-fidelity environments do not adversely impact the ability of the low-fidelity data to be used to predict high-fidelity performance.

This is important if we wish to train our model using low-fidelity data without quantitative adjustment.  If there was lower approximation efficiency between the low- and high-fidelity speed data, then quantitative methods would have to be leveraged in order to combine the data in a meaningful manner; previous theoretical studies investigated such methods \cite{Schlicht:2012} and when they should be leveraged.

\subsection{Impact of IVS Technology on Low-Fidelity Speed}

\begin{figure}[ht]
\vskip 0.2in
\begin{center}
\centerline{\includegraphics[width=.8\columnwidth]{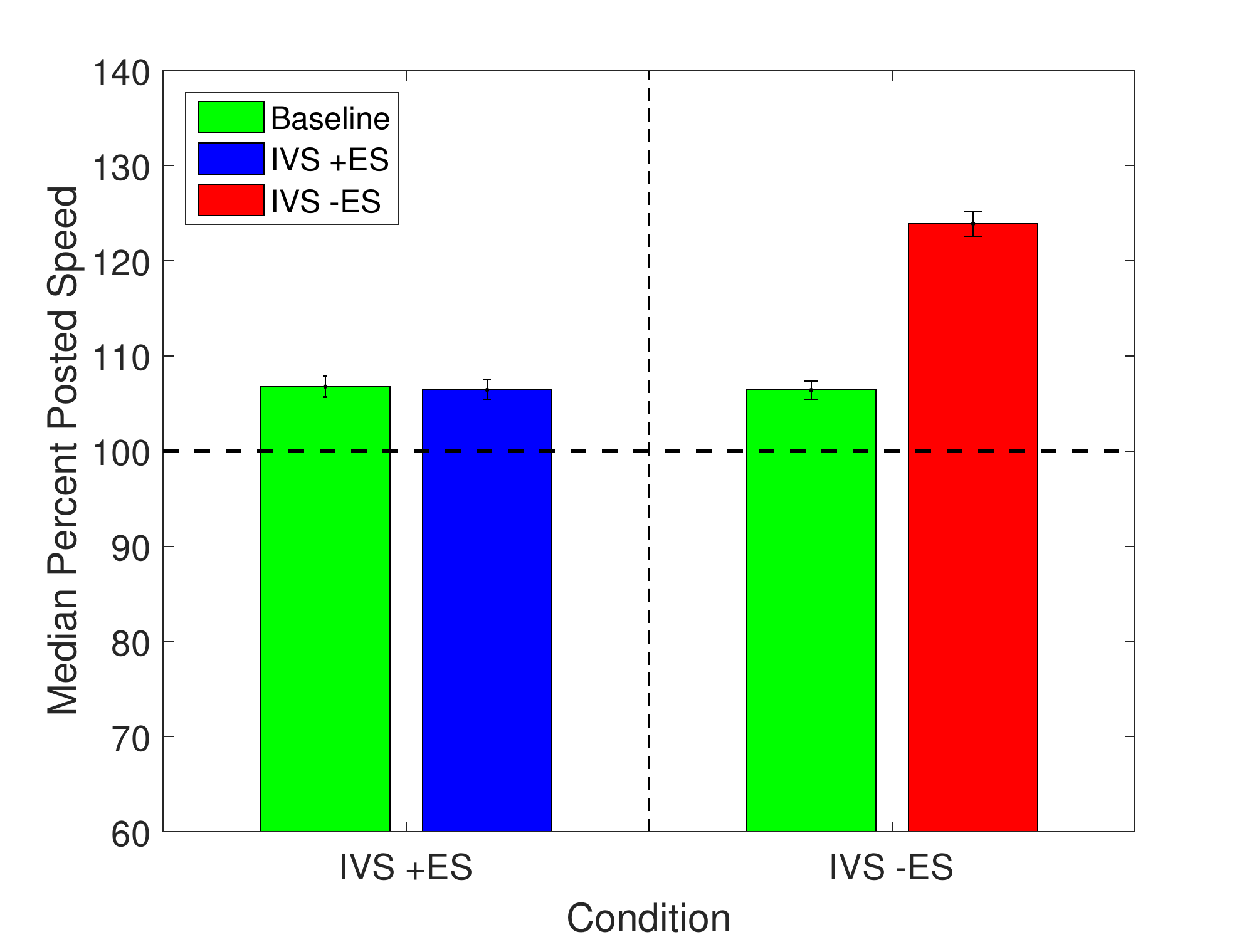}}
\caption{Median Low-fidelity speeding behavior across IVS technology conditions}
\label{fig:speed}
\end{center}
\vskip -0.2in
\end{figure}

Figure \ref{fig:speed} shows the median percent posted speed averaged across all participants $i$ and zones $z$ during low-fidelity data collection.  Speed was compared for the conditions when IVS information was presented in conjunction with roadside signs (IVS +ES) and conditions in which IVS information was presented in the absence of roadside sinage (IVS -ES).  We found that percent posted speed was relatively consistent across technology conditions for the baseline $(M=106.79, SE=1.10)$ and IVS +ES condition $(M=106.44, SE=1.04)$. However, there were differences in median percent posted speed across technology conditions for the baseline $(M=106.42, SE=.94)$ and IVS -ES $(M=123.90, SE=1.31)$ conditions.

To explore if these differences were significant, a mixed-factorial ANOVA was performed on the data. The mixed-factorial ANOVA found the observed differences in speed to be significant, as there were significant main-effects of both IVS condition $(F(1,636) = 141.41, p<.01)$ and technology conditions $(F(1,636) = 25.66, p<.01)$.  Moreover, there was a significant interaction between technology and IVS conditions $(F(1,636) = 121.04, p<.01)$, where those in the IVS -ES group displayed significantly greater speeds over posted values than those in the other conditions. Clearly, this interaction was driving the significant main-effects that were observed.

Although it is obvious that increased speed in the IVS -ES condition will result in increased property damage and injury severity in the event of a crash, it is unclear about the extent to which this is true. After all, if in-vehicle sign information were to replace external signs, it would presumably save money on infrastructure costs, \emph{so it is desirable to understand the balance of these two factors in order to make an informed decision about the relative utility of the IVS technology.}

In order to provide a proof-of-concept of how such an estimate could be established, we focused on how the observed increases in speed will impact the expected lives lost, in the event of different types of crashes.  The next section will overview the predictive model and risk analysis done to that end.

\section{Multifidelity Risk Estimation}
In order to estimate the risk involved with a system, performance needs to be evaluated across several conditions.  It is intractable to run human-in-the-loop experiments over all types of people and situations.  Moreover, it is impossible to leverage high-fidelity data when evaluating the risk associated with a system that has yet to be deployed. Therefore, it is desirable to develop a model that is able to \emph{predict} human performance across the technological and environmental conditions of interest. The model can be used in Monte-Carlo simulation to evaluate the risk associated with deploying different types of IVS technology.  This section will describe the model and assumptions used for our efforts to this end.
\begin{figure}[ht]
\vskip 0.2in
\begin{center}
\centerline{\includegraphics[width= .6\columnwidth]{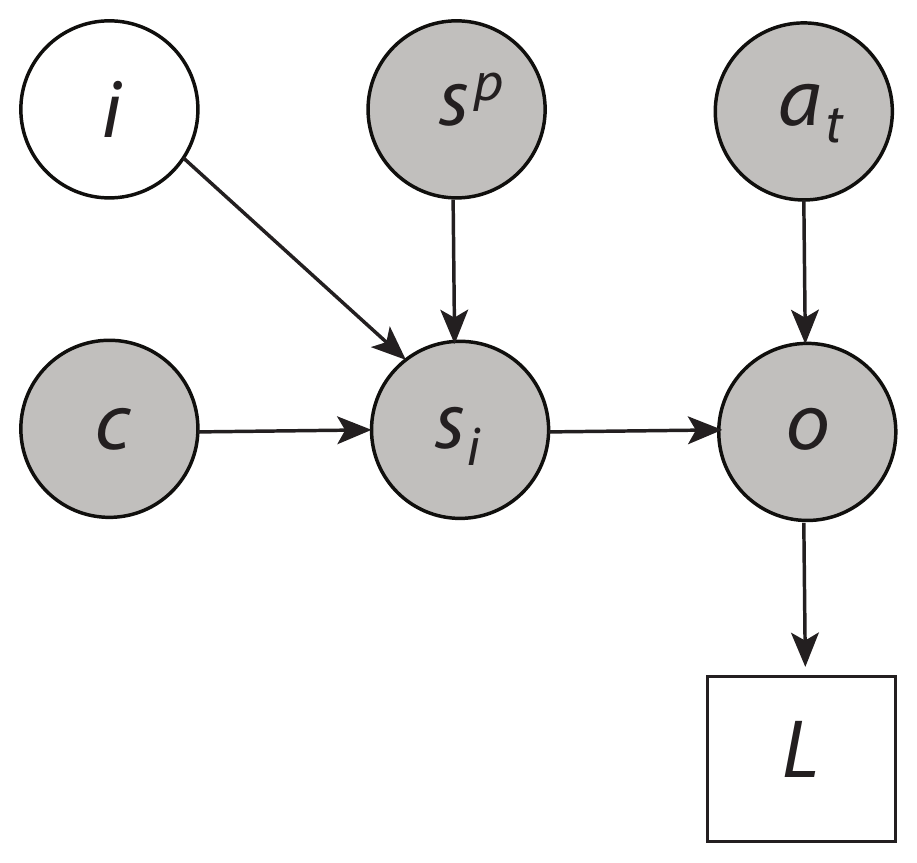}}
\caption{Bayesian network for IVS risk estimation}
\label{fig:bn}
\end{center}
\vskip -0.2in
\end{figure}
In the context of the current study, we would like to estimate the fatality risk associated with different IVS technology conditions ($c$).  Figure \ref{fig:bn} shows the Bayesian network used for the IVS Monte-Carlo simulation.  Shaded circles represent observable random variables, white circles represent unobservable random variables, squares represent the loss function associated with realizing the possible outcomes ($o$), and arrows represent causal relationships between variables.

Formally, this directed-acyclic graph (DAG) represents the following probabilistic relationship:

\begin{equation}
\label{eq:1}
\begin{aligned}
\mathbb{E}(V_c(o))  &= \sum_j L(o_j) \sum_{i\in I} p(o_j \mid a_t,s_{i})p(s_{i} \mid s^p, i,c)p(s^p)p(i)\\
&= \sum_j L(o_j) p(o_j \mid a_t,s)p(s \mid s^p,c)p(s^p)
\end{aligned}
\end{equation}

where $\mathbb{E}(V_c(o))$ is the expected-value associated with IVS condition $c$ across outcomes $o$.  $L(o_j)$ is the loss associated with realizing outcome $j$, and $p_c(o_j \mid s,a_t)$ is the probability of realizing outcome $j$, given the speed $s$ and accident type $t$. The conditional probability $p(s \mid s^p,c)$ represents the probability of observing speed $s$ given the posted speed $s^p$ and IVS condition $c$. This conditional probability was derived by marginalizing over individual participant factors, ($I$), which also impact the rates of observed speed. Finally, $p(s^p)$ is the marginal probability of posted speeds across a region (temporal or geographical) of interest.

In the general case, $L(o_j)$ could quantify property damage, injury severity and fatality rates.  However, for the purpose of the current study, we will utilize a loss-function that focuses on fatality rates.  More specifically, our loss function simply provides a unit reward for a positive outcome $L(o = 1) = +1$, and a negative unit penalty for a fatality $L(o = 0) = -1$.

Notice that in order to reliably estimate $p(s \mid s^p, c)$, it requires that we quantify the distribution of observed speeds \emph{in the real-world} across various speed zones ($s^p$), operating under different IVS technology conditions $c$.  This is intractable, as we can only use high-fidelity traffic data to estimate this distribution for the baseline condition $p(s \mid s^p, c = $ baseline technology$)$; the IVS conditions do not occur outside our low-fidelity simulation, so we need to develop a model that can help us \emph{predict} $p(s \mid s^p, c)$ for the IVS +ES and IVS -ES conditions.

Candidate predictive models were proposed in the order of increasing complexity (i.e., in the number of features and model parameters). It was discovered that a quadratic model seemed to be the best balance between model simplicity and predictive performance.  We converged on a very simple nonlinear model that uses only two features:
\begin{enumerate}
\item Speed in the baseline condition for each zone (i.e., posted speed zone, $z$), averaged across participants $s_z^b$. Essentially, this feature encodes average speeding behavior for a particular posted speed zone when no IVS is present.  Notice that this feature can be estimated via high-fidelity or low-fidelity data.
\item IVS condition indicator variable encoding the absence or presence of external signs $\delta_{c} \in {0,1}$, respectively.
\end{enumerate}

Specifically, the model takes-on the following form:

\begin{equation}
\label{eq:2}
s^{c}_z = w_0 + w_1*s_z^b + w_2*\delta_{c} + w_3*(s_z^b)^2
\end{equation}

\begin{figure}[ht]
\vskip 0.2in
\begin{center}
\centerline{\includegraphics[width= .8\columnwidth]{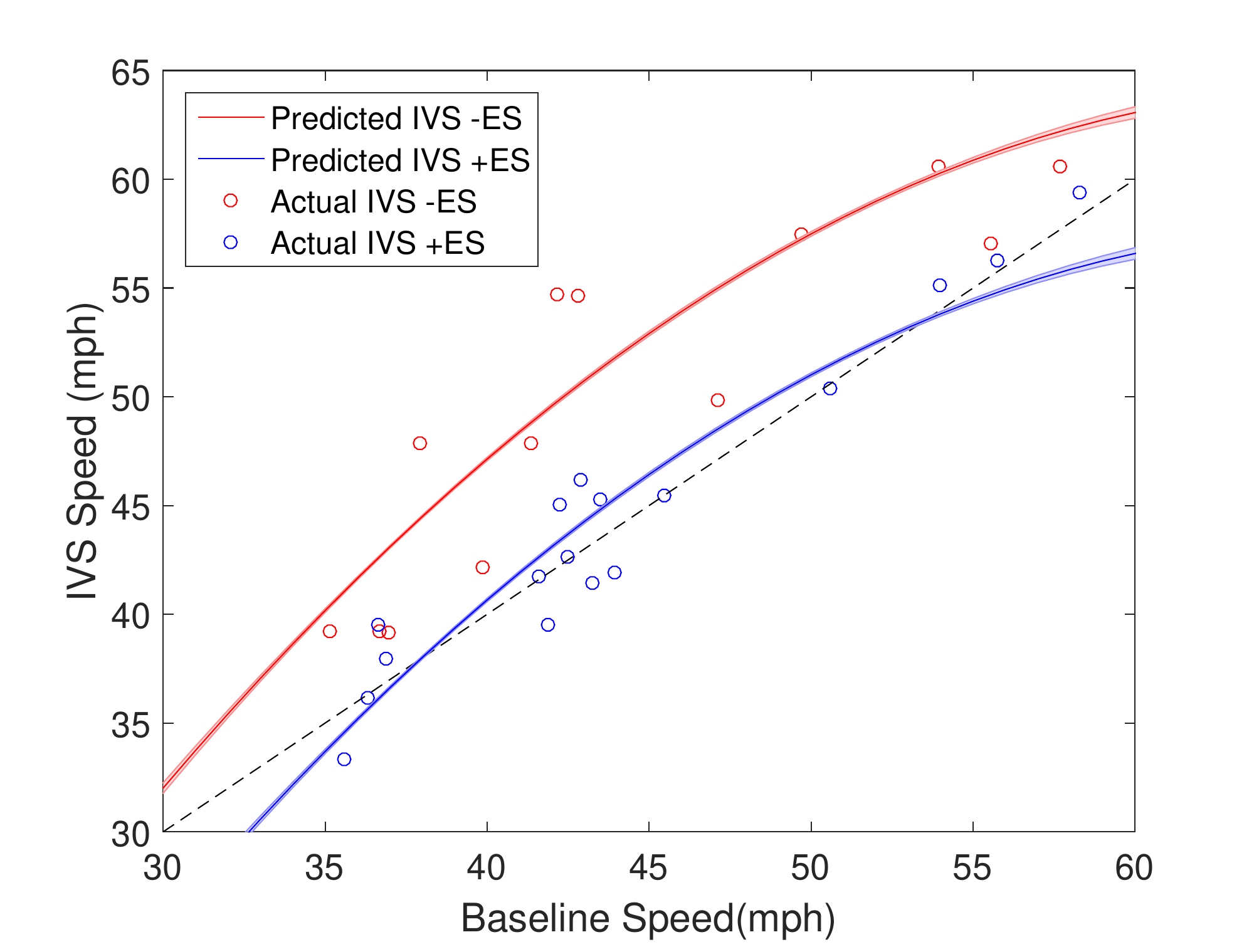}}
\caption{Training performance of model}
\label{fig:model}
\end{center}
\vskip -0.2in
\end{figure}

Figure \ref{fig:model} shows the model performance on training data across baseline speeds for each IVS condition.  The solid colored line shows the mean prediction across each of the 10 model weights $w$ estimated from the k-fold cross validation procedure, and the shaded regions represent $\pm$ 1 SEM.

\begin{figure}[ht]
\vskip 0.2in
\begin{center}
\centerline{\includegraphics[width= .8\columnwidth]{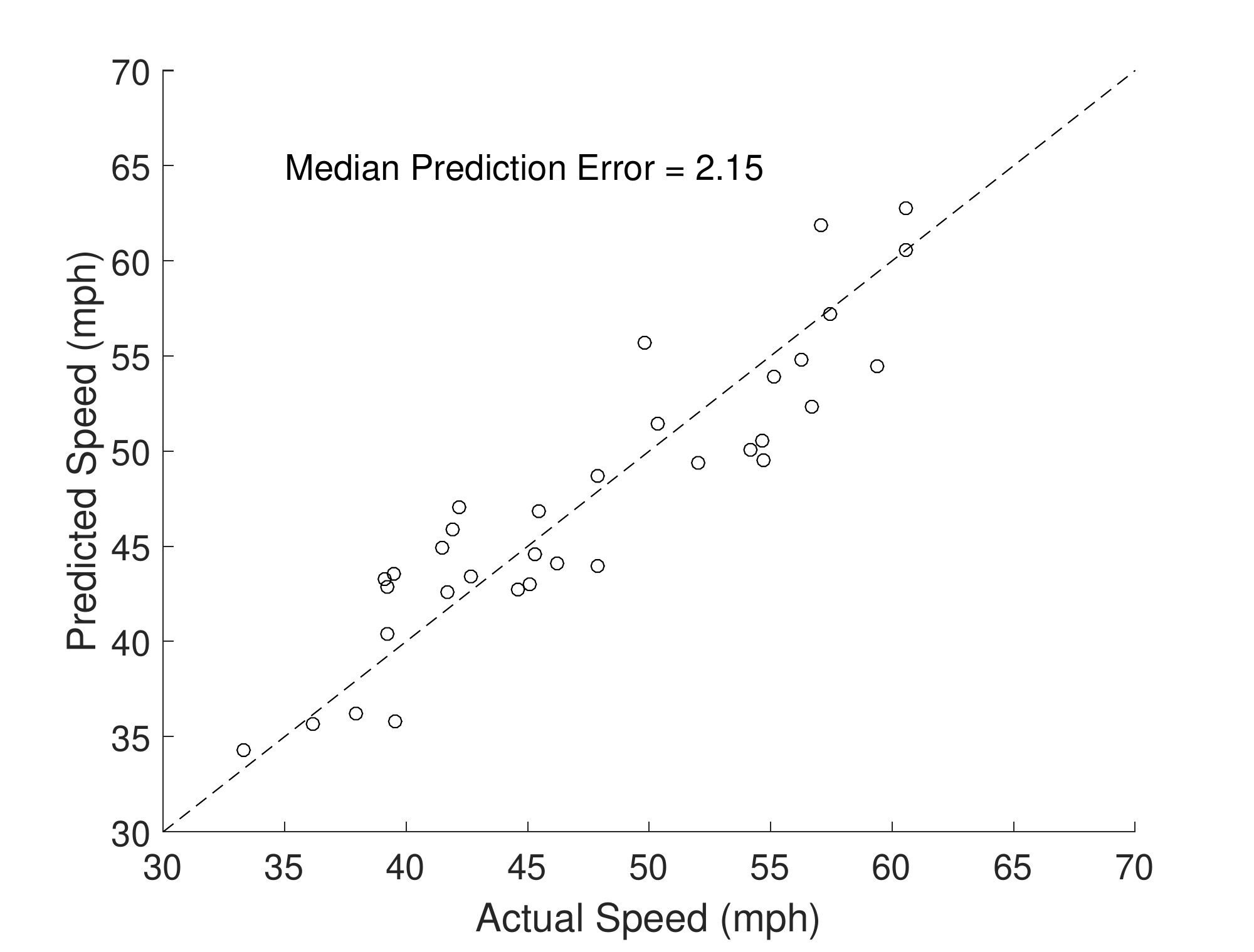}}
\caption{Predictive performance of model}
\label{fig:predperf}
\end{center}
\vskip -0.2in
\end{figure}

Figure \ref{fig:predperf} shows the predictive performance of the model. Predictive performance was evaluated using k-fold cross-validation, where the data set is partitioned into data that is used to estimate model parameters (i.e., training data) and data that is used to evaluate predictive performance (i.e., test data).  This partitioning procedure is performed $k=10$ times across the data and the results of the models ability to predict the test data is shown on Figure \ref{fig:predperf}. If the model resulted in perfect prediction, all the data would fall on the dashed-line.  As the figure shows, the quadratic model defined in Equation \ref{eq:2} is able to predict speed in each IVS condition from baseline data, with a median prediction error of $\pm$ 2.2 mph.

\begin{figure}[ht]
\vskip 0.2in
\begin{center}
\centerline{\includegraphics[width= .8 \columnwidth]{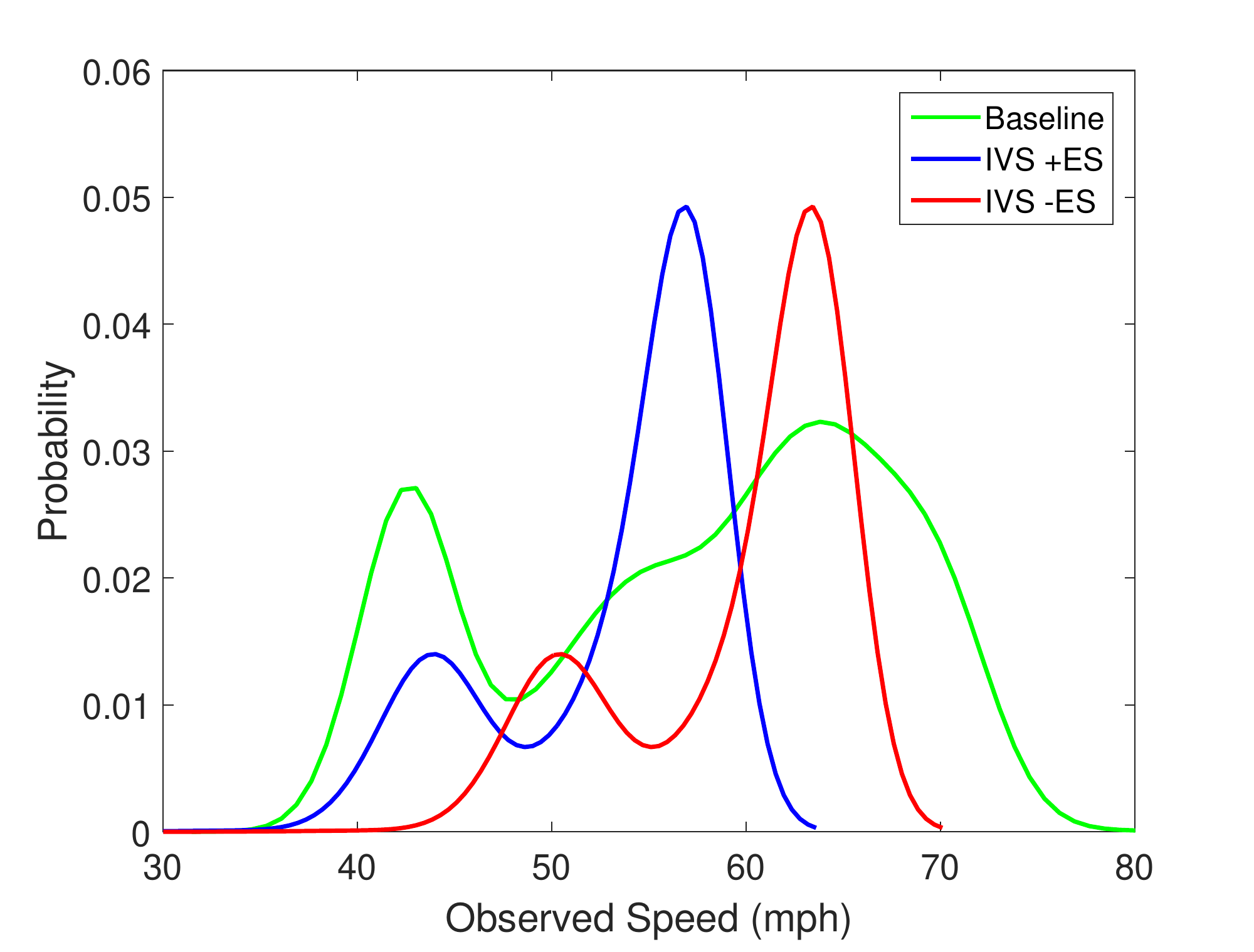}}
\caption{Sampling distributions used in risk estimation}
\label{fig:sampling}
\end{center}
\vskip -0.2in
\end{figure}

In order to estimate $p(s^{c}_z \mid s_z, c = $ baseline technology$)$, we used data from a 2014 MnDOT ATR report. Speed data was provided for four different speed zones (40, 50, 55 and 60 mph), and the frequency of the observed speeds for a given zone were binned using bin sizes of 5mph.  Since we need to estimate the average baseline speed to use as a feature for our predictive model, we computed the weighted average speed observed in each zone once an hour, and the resulting distribution of those hourly speeds was used to estimate $p(s^{c}_z \mid s_z, c = $ baseline technology$)$.  Using this high-fidelity speed data, we performed kernel density estimation on the average baseline speed distributions for each zone ($z$).  Then, average speeds were sampled from the baseline distribution (N = 10,000), and the corresponding speeds for the different IVS technology conditions were predicted by using the baseline samples as inputs into Equation \ref{eq:2}. Now that we have predicted observed speeds $s^{c}_z$ for each IVS condition ($c$) and posted speed zone ($z$), we need to estimate $p(s^{c}_z \mid s_z,c)$.  This was also accomplished by using kernel density estimation with a kernel width of two.  Figure \ref{fig:sampling} shows the results of estimating distributions associated with each IVS technology condition , marginalizing over all posted speed zones $p(s^{c} \mid c) = \sum_z p(s^{c}_z \mid s_z,c)$.

\begin{figure}[ht]
\vskip 0.2in
\begin{center}
\centerline{\includegraphics[width= .8\columnwidth]{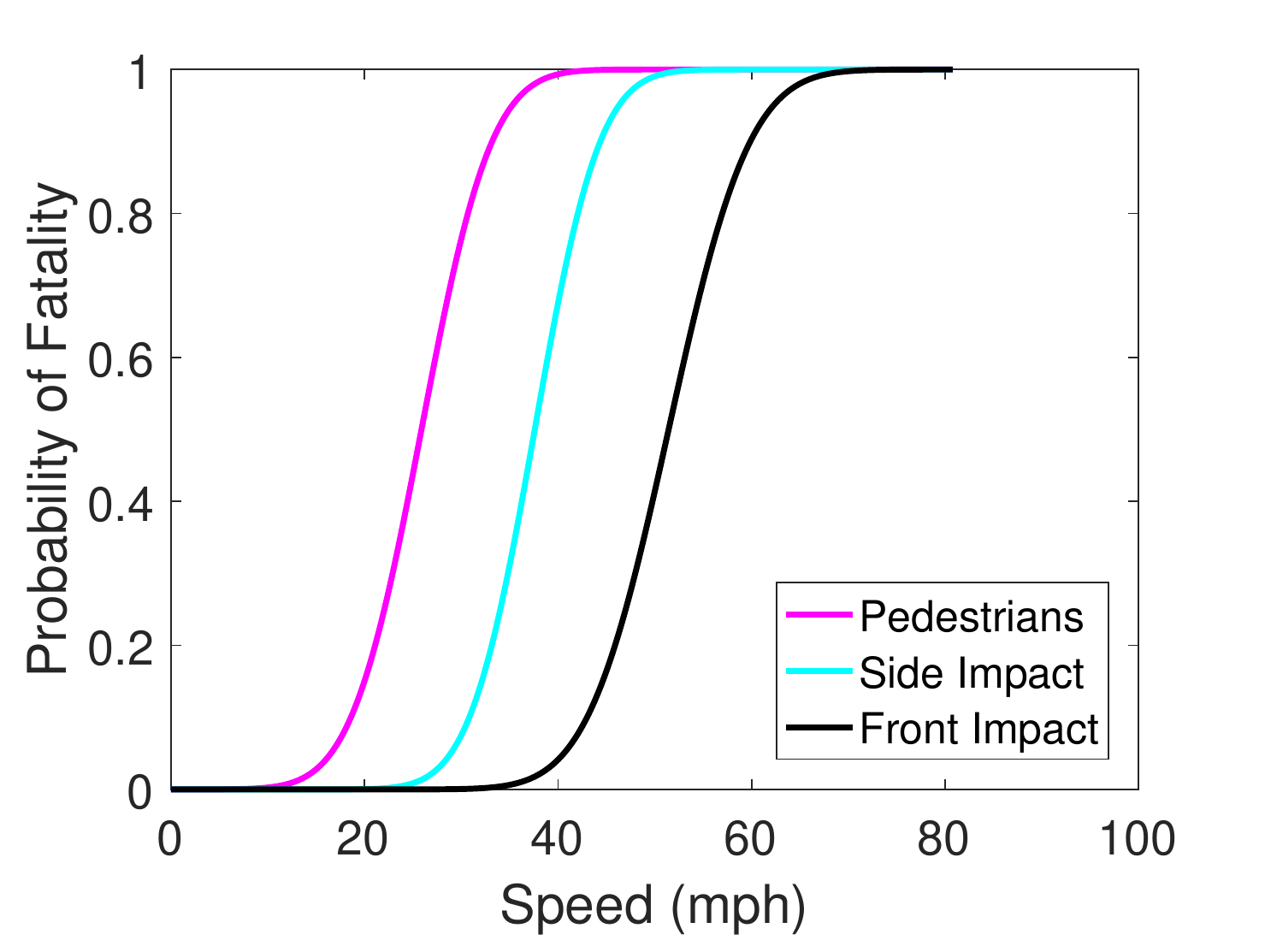}}
\caption{Probability of fatality across accident types}
\label{fig:probit}
\end{center}
\vskip -0.2in
\end{figure}

The distribution $p_c(o_j = 1 \mid a_t = t,s)$ needs to be estimated from high-fidelity data, as we did not experience any crashes in low-fidelity simulation.  This was accomplished by leveraging high-fidelity traffic data from an international source.\footnote{\href{http://www.audit.vic.gov.au/publications/2011-12/20110831-Road-Safety-Cameras/20110831-Road-Safety-Cameras.html}{Data adapted from Victorian Government Online Report, Figure 2C}} Probit regression was utilized to recreate the fatality curves for each of the three crash types (i.e., pedestrian, side-impact, front-impact). More specifically, the probability of fatality for each speed and crash type was given by $p_c(o = 0 \mid a_t=t,s)$, and is depicted in Figure \ref{fig:probit}.  The probability of surviving a crash $p_c(o = 1 \mid a_t=t,s)$ is simply $1 - p_c(o = 0 \mid a_t=t,s)$. Although the high-fidelity data are from an international source, the results should be valid across a wide range of situations, as the data should not be strongly correlated with geographical factors.  However, the marginal probability of crash type $p(a_t)$ may vary across regions, but this will not impact our results, as we are computing risk separately across crash types, as described below.

Finally, in order to simplify our expected value calculations, we only compute the expected-value associated with each IVS condition in the event of an crash. Moreover, we compute this separately for each crash type, which simplifies Equation \ref{eq:1} to:
\begin{equation}
\label{eq:3}
\mathbb{E}(V^t_c(o))  = \sum_j L(o_j)p(o_j \mid a_t=t,s)p(s \mid s^p,c)p(s^p)
\end{equation}

\begin{figure}[ht]
\vskip 0.2in
\begin{center}
\centerline{\includegraphics[width= \columnwidth]{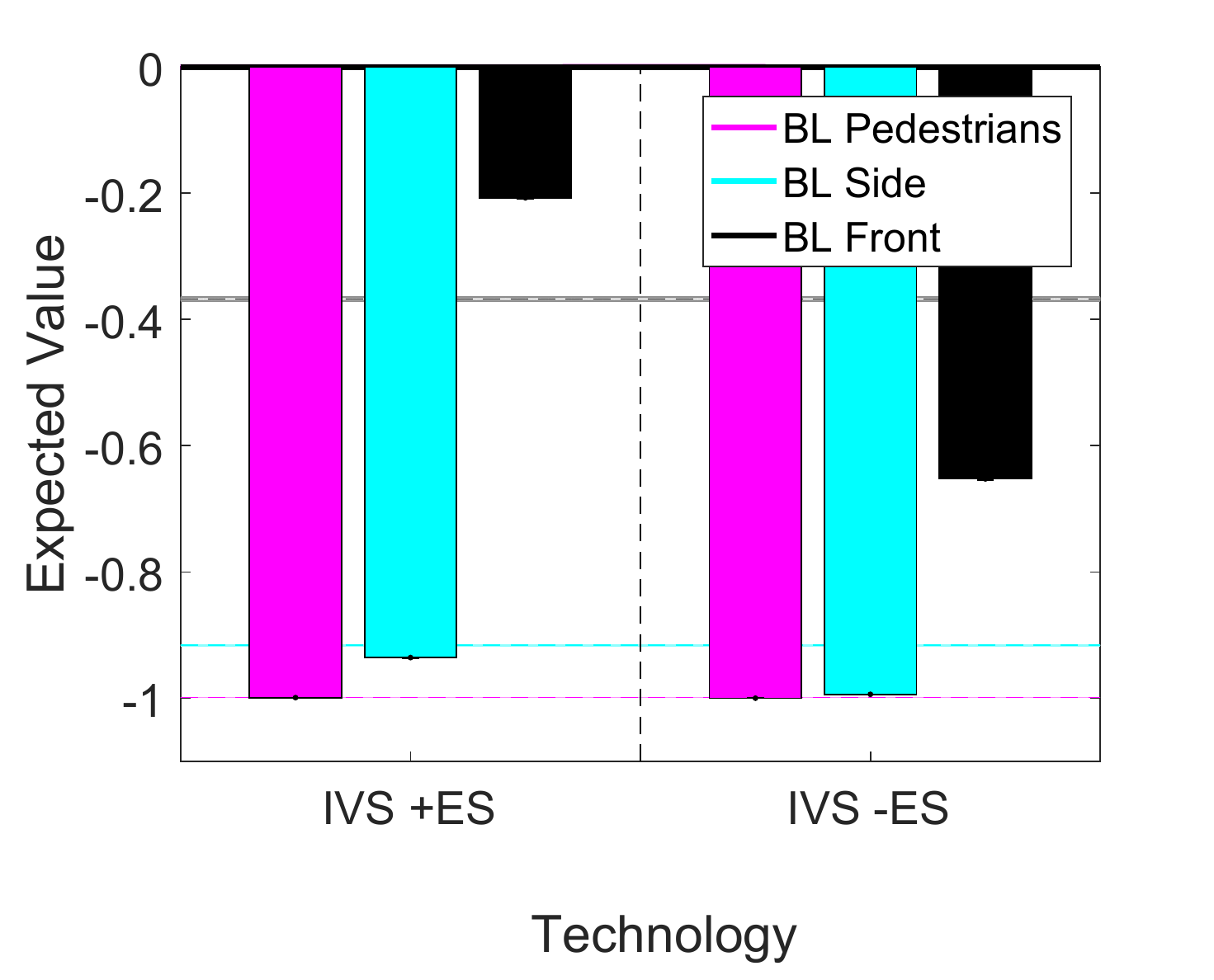}}
\caption{Average risk across technology conditions and accident types}
\label{fig:risk}
\end{center}
\vskip -0.2in
\end{figure}

Equation \ref{eq:3} is what was utilized to evaluate risk for the IVS conditions in this study.  Leveraging the loss function and probability estimates described above, we were able to perform Monte-Carlo simulation (Algorithm \ref{algo}) to produce the IVS risk estimates found in Figure \ref{fig:risk}.

\begin{algorithm}
\caption{IVS Monte-Carlo Risk Simulation}\label{algo}
\begin{algorithmic}[1]
\FOR {$n \in N_{simulations}$}
\STATE $s_n^p \gets sample \sim p(s^p)$
\FOR {$c \in C$}
\STATE $s_c \gets sample \sim p(s \mid s_n^p,c)$
\FOR {$t \in T$}
\STATE $p^t_c \gets  p(o_j = 0 \mid a_t=t, s_c)$
\STATE $ev^t_{c,n} \gets  (p^t_c \times L(o=0)) + ((1-p^t_c) \times L(o=1))$
\ENDFOR
\ENDFOR
\ENDFOR
\STATE $\mathbb{E}(V_c^t(o)) = \frac{1}{N} \times \sum_n ev^t_{c,n}$
\end{algorithmic}
\end{algorithm}

Figure \ref{fig:risk} shows the average EV associated with different crash types across IVS conditions.  In this formulation, expected-value can be intuited as the expected number of lives lost in the event of an unmitigated crash of type $t$.  Expected-value, in this case, ranges between -1 and +1.  The greater the expected-value, the greater the safety of the system.  In this respect, we are able to evaluate the safety associated the IVS technology, and the goal of designing an IVS system should be to maximize the EV associated with the technology, which corresponds to minimizing the expected risk.

To compare relative safety of the IVS technology, we computed the EV associated with the baseline condition (Figure \ref{fig:risk}, horizontal colored lines). It was found that EV in the baseline condition for an crash involving a pedestrian was the least safe (M = -.99, SE = .01), and EV associated with a side- and (M = -.92, SE = .01) front-impact was also negative (M = -.37, SE = ,01). The negative estimated risk is what would be expected from the fatality curves depicted in Figure \ref{fig:probit}, and the speed distributions that were provided by MnDOT and used for this risk simulation (Figure \ref{fig:sampling}). For example, the near negative-one EV associated with crashes involving a pedestrian results from the fact that almost all the observed speeds for which we had high-fidelity data are greater than the speed associated with $p(o_j = 0 \mid a_t=t, s_c) = 1$.

Expected-value for the IVS +ES condition demonstrates that the technology actually improved safety relative to baseline conditions in the case of an crash involving a front impact (M = -0.21, SE = .01).  However, for crashes involving side-impact (M = -.94, SE = .01) and pedestrians (M = -.99, SE = .01), EV was comparable to levels in the baseline condition, suggesting that the IVS technology minimally impacts the safety associated with those crash types.

Finally, expected-value for the IVS condition without external signs (-ES) demonstrates that the technology significantly decreased safety relative to baseline conditions.  This was true across crashes involving pedestrians (M = -1.00, SE = .01), side- (M = -.99, SE = .01) and front-impact (M = -.65, SE = .02). Clearly, the increases in speed observed (Figure \ref{fig:speed}) in the IVS -ES condition adversely impacts crash safety expectation.

\section{Conclusions}
This study proposed a multifidelity method to quantitatively evaluate the risk associated with transportation technology, \emph{prior to deployment}.  As proof-of-concept, we estimated the risk associated with in-vehicle sign (IVS) technology and found that relying on IVS information alone led to a significant increase in expected fatalities for crashes involving front- and side-impact, relative to status quo conditions (Figure \ref{fig:risk}).  The increased risk was the result of increases in speed under IVS technology conditions where external signs were absent (Figure \ref{fig:speed}).

However, the analysis also discovered that presenting IVS information led to decreased risk when paired with external signs (Figure \ref{fig:risk}), relative to conditions where external roadside signs were only present.  More specifically, risk analysis suggests that there are fewer expected fatalities involving front-impact collisions.  Interestingly, in this case, the reduction in risk was not the result from decreased average speeds (Figure \ref{fig:speed}), but rather from the properties of the  \emph{distribution} of speeds predicted under this IVS condition (Figure \ref{fig:sampling}).  Therefore, technology evaluation that exclusively relied on behavioral speed averages from low-fidelity simulation would have been remiss.

This study presented a novel multifidelity method for estimating the risk associated with transportation technology prior to deployment. Future extensions could include the addition of monetary concerns; both direct (e.g., cost of collision repair) and comprehensive (e.g., wages lost) costs could be incorporated into the loss function.  By doing this, risk would better reflect concerns of individuals who are responsible for evaluating transportation technology.

Future work could include updating the fatality probability curves with data from domestic sources.  Moreover, the fatality probability curves could be conditioned on other factors that are known to change with modifications in transportation technology.  Ideally, our probability curves would also include a model for how variability in speed impacts factors that are being considered by the loss function (e.g., fatalities, cost).

Overall, these results provide a proof-of-concept for how multifidelity models can be leveraged to estimate the risk associated with transportation technology, \emph{prior to deployment}. This allows for candidate technologies to be evaluated at the stage of conception, and enables a mechanism for only the safest and most effective technology to be developed and released.

\section*{Acknowledgements}

This analysis was funded Minnesota Local Road Research Board and performed under MnDOT Contract Number 99008, Work Order Number 145. A special thanks to Victor Lund, project Technical Liason and Traffic Engineer of St. Louis County, MN. The authors would like to thank Janet Creaser, Michael Manser, and Peter Easterlund of the HumanFIRST Program, in addition to Alec Gorjestani of the Intelligent Vehicles Laboratory for their assistance during the experimental design and low-fidelity data collection portion of this project. Moreover, we would like to thank Derek Leuer, Nathan Drews and Eric DeVoe of MnDOT for providing some of the high-fidelity traffic data used for our risk analysis.

\nocite{langley00}

\bibliography{schlicht_multifidelity}
\bibliographystyle{icml2017}

\end{document}